\documentclass[12pt, a4paper]{article}

\usepackage{amsmath, amssymb}
\usepackage{geometry}
\usepackage{hyperref}
\usepackage{setspace}
\usepackage{parskip}
\usepackage{titlesec}
\usepackage[numbers]{natbib}
\usepackage{tikz}
\usepackage{pgfplots}
\pgfplotsset{compat=1.18}
\usepackage{float}       
\usepackage{caption}    
\usepackage{graphicx}    
\usepackage{subcaption}
\usetikzlibrary{patterns}

\hypersetup{colorlinks=true, linkcolor=blue, citecolor=blue, urlcolor=blue}

\titleformat{\section}{\normalfont\large\bfseries}{\thesection.}{0.5em}{}
\titleformat{\subsection}{\normalfont\normalsize\bfseries}{\thesubsection}{0.5em}{}

\begin{document}

\begin{center}
  {\LARGE\bfseries Topology-Informed Survival Analysis of Breast Cancer Patients Using the  Mapper Algorithm}\\[14pt]
  {\large Emmanuel Kibisi$^{1}$, Olakunle Abawonse$^{1}$,
         and Donald Woukeng$^{1}$}\\[6pt]
  {\normalsize $^{1}$African Institute for Mathematical Sciences (AIMS), Rwanda}\\[4pt]
  {\normalsize \texttt{emmanuel.kibisi@aims.ac.rw}}\\[0.1pt]
  {\normalsize \texttt{olakunle.abawonse@aims.ac.rw}}\\[0.1pt]
  {\normalsize \texttt{donald.woukeng@aims.ac.rw}}\\[14pt]
\end{center}

\hrule
\vspace{10pt}

% ABSTRACT
\noindent\textbf{Abstract.}\;
This study applied a mathematical tool from Topological Data Analysis (TDA), called the Mapper algorithm, to gene expression data from more than 1,000 TCGA-BRCA patients to identify hidden molecular patterns associated with survival. Patients located near high-risk regions of the network showed significantly poorer survival, and highly proliferative gene expression patterns were associated with worse outcomes overall, although treatment narrowed this survival
gap across proliferation groups. The analysis further uncovered patients whose survival outcomes were inconsistent with their expected clinical behavior, including a subgroup of Basal-like patients with unexpectedly favorable outcomes linked to a distinct, more treatment-responsive gene signature, revealing molecular programs missed by traditional classification methods. Validation through training and testing on unseen patients confirmed that topology-derived risk groups remained significantly associated with survival after adjusting for age, tumor stage, and treatment, demonstrating that the geometric structure of gene expression data contains clinically meaningful prognostic information beyond traditional breast cancer classification methods.\\

\vspace{6pt}
\noindent\textbf{Keywords:} Topological Data Analysis; Mapper Algorithm;
Breast Cancer; Gene Expression; Survival Analysis; Risk Stratification;
Discordant Survivors; TCGA-BRCA; Cox Proportional Hazards;
Kaplan--Meier; PAM50.
\vspace{10pt}
\hrule
\vspace{14pt}

\section{Introduction}
Breast cancer is now understood not to be a single disease but a collection of diseases which behave differently, with almost 2.3 million new diagnoses each year worldwide~\cite{sung2021global}. As a result, the clinical outcomes of patients presenting with the disease may vary significantly. In recent years, PAM50 classification has been used extensively to classify these variants using their specific molecular gene expression characteristics into intrinsic subtypes: Luminal A, Luminal B, Basal-like, and HER2-enriched, the last of which is driven by amplification of human epidermal growth factor receptor 2 (HER2)~\cite{parker2009supervised}. To refine these classifications, recent broad-scale genomic efforts have leveraged multi-algorithm machine learning consensus models to establish robust prognostic signatures capable of capturing therapeutic benefit across heterogeneous patient cohorts~\cite{Cheng2026}, while deep learning and transfer learning frameworks have been deployed to isolate survival subtypes from high-dimensional single-cell cellular interaction profiles~\cite{Yadav2023}. However, forcing these profiles into rigid categories results in misclassification  between borderline subtypes, such as Luminal A and B~\cite{ades2014Luminal} because these tumors exist on a continuous biological spectrum rather than within discrete boundaries. Recent literature increasingly turns to Topological Data Analysis (TDA), specifically the Mapper algorithm, to capture the high-dimensional geometric shapes and global connectivity of the underlying expression data. Nicolau et al.~\cite{nicolau2011topology} pioneered the application of Mapper to omics data through the Progression Analysis of Disease (PAD) framework which used the approach of transforming transcriptomics data utilizing Disease-Specific Genomic Analysis (DSGA) with Mapper technology to find patient subsets suffering from breast cancer. Building on this, Rostami et al.~\cite{rostami2025tda} demonstrated that 
the Mapper algorithm~\cite{singh2007topological} extracts the coordinate-free topological shape of complex gene expression profiles by constructing a simplified network in which nodes represent clusters of biologically similar patients and edges connect clusters sharing common individuals. They extended this by simultaneously constructing a feature network of gene clusters alongside the sample network, whereby this dual-space analysis not only delineated known breast cancer subtypes but also uncovered a novel Luminal B subgroup defined by distinct HER2 expression levels.

While the original dual-network characterization proposed by Rostami et al. successfully maps continuous gene-to-patient geometries to identify a hidden sub-population within clinically designated Luminal B disease, it stops short of evaluating how these topological structures correlate with actual clinical outcomes~\cite{rostami2025tda}. Similarly, contemporary clinical frameworks like Progression Analysis of Disease with Survival (PAD-S) by SurvMap, introduced by Jaume Forés-Martos et al.~\cite{fores2022progression}, embed outcome markers directly into their model lenses to map prognostic subgroups but lack rigorous independent validation, opting instead to partition and re-test the same primary study cohorts. This leaves a critical generalizability gap in topological oncology, where models frequently reveal striking qualitative shapes in high-dimensional datasets but remain unverified against standard clinical risk metrics or out-of-sample prediction pipelines~\cite{Loughrey2021}. Other topological frameworks like Survival Topological Representation Analysis of Diagrams (STRAND) attempt to bypass qualitative abstractions by vectorizing persistence diagrams directly as time-to-event functions, yet they discard standard spatial traceability, leaving clinicians unable to trace individual patient trajectories back to tangible node graphs~\cite{Murris2026}. To address these limitations, this study extends the topological framework of Rostami et al.~\cite{rostami2025tda} by performing the first systematic  survival analysis on a dual-Mapper foundation, where each patient's network position is linked to clinical outcome, discordant survivor subgroups are identified at subtype boundaries, and topology-derived risk tiers are validated against established clinical covariates on unseen patients using  Cox proportional hazards modeling. The main computational contribution is an interpretable workflow for deriving, transferring, and validating Mapper-based survival risk tiers from transcriptomic data.
The paper first introduces the mathematical foundations of the Mapper algorithm, followed by the survival analysis framework, and concludes with the results and their clinical interpretation.

\subsection{Aims and Objectives}

The primary aim of this study is to investigate the relationship between
topological network position and survival outcomes in TCGA-BRCA patients.
Specifically, the study seeks to:
\begin{enumerate}
  \item Reproduce the Mapper-based patient and gene co-expression networks
        from Rostami and co-workers~\cite{rostami2025tda}, validating pipeline
        reproducibility on the TCGA-BRCA cohort.
  \item Perform node-level survival analysis using Kaplan--Meier curves,
        log-rank tests, and Cox proportional hazards models to identify
        high-risk topological landmarks and characterise the survival gradient
        across the network.
  \item Identify discordant survivor subgroups at subtype-boundary nodes and uncover their underlying gene expression programs.
  \item Validate the generalizability of topology-derived risk tiers to
        unseen patients through a train-test framework, testing whether
        topological risk stratification retains independent prognostic value
        after adjusting for all established clinical covariates.
\end{enumerate}

\section{Preliminaries}

\subsection{The Mapper Algorithm}
The Mapper algorithm~\cite{chazal2021introduction} is a method from topological data analysis that reduces the complexity of high-dimensional data while preserving its underlying non-linear structure. Given a point cloud $X \subset \mathbb{R}^m$, the algorithm produces a graph $G = (V, E)$ that summarises the topology of $X$ through four sequential steps: data standardization, lens projection, cover construction, and local clustering. Each vertex $v \in V$ corresponds to a cluster of points in $X$, and an edge $(v_i, v_j) \in E$ exists whenever the corresponding clusters share at least one data point.

\subsubsection{Data standardization}

Let $X \in \mathbb{R}^{n \times m}$ be a data matrix with $n$ observations and $m$ features. Each feature $j$ is standardised as
$$
z_{ij} = \frac{x_{ij}-\mu_j}{\sigma_j},
$$
where $x_{ij}$ is the raw value of the $i$-th observation for the $j$-th variable, $\mu_j$ is the mean and $\sigma_j$ is the standard deviation of the $j$-th feature across all observations.

\subsubsection{The Lens (Filter) Function}
A lens function $f: X \to \mathbb{R}^d$, where mostly $d \in \{1,2\}$, is a continuous map that projects a high-dimensional dataset $X$ onto a lower-dimensional space while preserving its core geometric structure. Choices of $f$ include density estimators, eccentricity metrics, and the graph Laplacian, with 
each choice distinctly shaping the final topological output~\cite{singh2007topological}. In this study, we employ $L_1$ eccentricity, which quantifies the mean 
deviation of an observation from the rest of the dataset. Formally, for the standardised matrix $Z \in \mathbb{R}^{n \times m}$, the $L_1$ (Manhattan) distance between observations $i$ and $k$ is
$$
d_1(Z_i,Z_k) = \sum_{j=1}^{m}|z_{ij}-z_{kj}|,
$$
and the $L_1$ eccentricity of the $i$-th point is defined as
$$
e_i = \frac{1}{n-1}\sum_{k \neq i}d_1(Z_i,Z_k),
$$
so that observations lying far from the bulk of the data attain large eccentricity values, making $e_i$ an effective lens for separating peripheral from central data points.

\subsubsection{Covers and the Nerve Construction}

Having defined the lens function $f: X \to \mathbb{R}$, the image $f(X)$ is covered by a finite collection of overlapping open intervals $\mathcal{U} = \{U_i\}_{i \in I}$ satisfying $f(X) \subseteq \bigcup_{i \in I} U_i,$ where each $U_i$ is parameterised by a fixed length $\eta > 0$ and overlap fraction $p \in (0,1)$.  The preimage $f^{-1}(U_i) = \{x \in X : f(x) \in U_i\}$ then defines the  subset of observations assigned to the $i$-th cover element. A clustering  algorithm (e.g., DBSCAN~\cite{ester1996density}) is then applied  independently within each preimage $f^{-1}(U_i)$, identifying $\ell(i)$ disjoint clusters $C_{i1}, C_{i2}, \ldots, C_{i\ell(i)}$; a single  cover element may yield several clusters whenever its preimage contains  disconnected regions of $X$. The Mapper graph $G=(V,E)$ is then the  $1$-skeleton of the nerve of this refined, clustered cover: the vertex set is the collection of all clusters, $V = \{C_{ij} : i \in I,\ j = 1, \ldots, \ell(i)\},$ and an edge exists between $C_{ia}$ and $C_{jb}$ whenever $C_{ia} \cap C_{jb} \neq \emptyset,$ that is, whenever the two clusters share at least one observation. Vertices  are therefore clusters, not cover elements: overlap between $U_i$ and $U_j$ 
makes an edge \emph{possible} but not guaranteed, since two clusters can 
arise from overlapping cover elements without themselves sharing any point.

\begin{figure}[H]
\centering
\begin{subfigure}[b]{0.35\textwidth}
\centering
\begin{tikzpicture}[scale=0.75]
  
  \path[use as bounding box] (-0.5,-3.9) rectangle (4.5,3.6);
  \draw[->, thick] (-0.5,-3.0) -- (4.5,-3.0);
  \foreach \x in {0,1,2,3,4} {
    \draw (\x,-2.9) -- (\x,-3.1) node[below] {\tiny \x};
  }
  \node[below] at (2,-3.55) {\scriptsize Lens $f(x,y)=x$};
  \draw[dashed, gray!80, fill=gray!5] (0,-2.5) rectangle (2,2.5);
  \draw[dashed, gray!80, fill=gray!10, fill opacity=0.5] (1,-2.5) rectangle (3,2.5);
  \draw[dashed, gray!80, fill=gray!5] (2,-2.5) rectangle (4,2.5);
  \fill[pattern=north east lines, pattern color=gray!60] (1,-2.5) rectangle (2,2.5);
  \fill[pattern=north east lines, pattern color=gray!60] (2,-2.5) rectangle (3,2.5);
  \node[above] at (2,2.5) {\scriptsize $U_2=(1,3)$};
  \node[left] at (1.2,2.0) {\tiny $U_1=(0,2)$};
  \node[right] at (3.0,2.0) {\tiny $U_3=(2,4)$};
  \foreach \a/\dx/\dy in {0/0.03/-0.03,15/-0.04/0.02,30/0.03/0.04,
    45/-0.03/-0.03,60/0.04/0.03,75/-0.03/0.04,90/0.03/-0.03,
    105/-0.04/-0.03,120/0.03/0.04,135/-0.03/0.03,150/0.04/-0.03,
    165/-0.03/0.03,180/0.03/-0.03,195/-0.04/0.02,210/0.03/0.04,
    225/-0.03/-0.03,240/0.04/0.03,255/-0.03/0.04,270/0.03/-0.03,
    285/-0.04/-0.03,300/0.03/0.04,315/-0.03/0.03,330/0.04/-0.03,345/-0.03/0.03}
    \fill[black!50] ({2 + 2.0*cos(\a)+\dx},{2.0*sin(\a)+\dy}) circle (0.045);
  \draw[teal, line width=2.5pt] ({2+2.1*cos(120)}, {2.1*sin(120)}) arc (120:240:2.1);
  \node[teal, left] at (0.1, 0) {\textbf{\scriptsize $C_{11}$}};
  \draw[violet, line width=2.5pt] ({2+2.1*cos(-60)}, {2.1*sin(-60)}) arc (-60:60:2.1);
  \node[violet, right] at (3.9, 0) {\textbf{\scriptsize $C_{31}$}};
  \draw[orange, line width=2.5pt] ({2+2.1*cos(60)}, {2.1*sin(60)}) arc (60:120:2.1);
  \node[orange, above] at (2, 2.12) {\textbf{\scriptsize $C_{21}$}};
  \draw[red!70!black, line width=2.5pt] ({2+2.1*cos(240)}, {2.1*sin(240)}) arc (240:300:2.1);
  \node[red!70!black, below] at (2, -2.12) {\textbf{\scriptsize $C_{22}$}};
  \node[black!80] at (0.5,-1.8) {\scriptsize $X$};
\end{tikzpicture}
\caption{Cover and clusters}
\label{fig:cover-clusters}
\end{subfigure}
\hfill
\begin{subfigure}[b]{0.28\textwidth}
\centering
\begin{tikzpicture}[scale=0.75, every node/.style={draw, circle, minimum size=0.8cm, inner sep=0pt, font=\scriptsize}]
  \path[use as bounding box] (-1.5,-3.9) rectangle (1.5,3.6);
  \node[fill=orange!20] (C21) at (0,1.5) {$C_{21}$};
  \node[fill=teal!20]   (C11) at (-1.2,0) {$C_{11}$};
  \node[fill=violet!20] (C31) at (1.2,0) {$C_{31}$};
  \node[fill=red!20]    (C22) at (0,-1.5) {$C_{22}$};
  \draw[thick] (C11)--(C21) (C21)--(C31) (C31)--(C22) (C22)--(C11);
\end{tikzpicture}
\caption{Mapper graph}
\label{fig:mapper-graph}
\end{subfigure}
\hfill
\begin{subfigure}[b]{0.32\textwidth}
\centering
\begin{tikzpicture}[scale=0.75, every node/.style={draw, rounded corners, minimum width=1.6cm, minimum height=0.6cm, font=\scriptsize}]
  \path[use as bounding box] (-1.5,-3.9) rectangle (1.5,3.6);
  \node (U1) at (0,1.3)  {$U_1 = (0,2)$};
  \node (U2) at (0,0)    {$U_2 = (1,3)$};
  \node (U3) at (0,-1.3) {$U_3 = (2,4)$};
  \draw[thick, ->] (U1)--(U2);
  \draw[thick, ->] (U2)--(U3);
\end{tikzpicture}
\caption{Naive cover path}
\label{fig:naive-path}
\end{subfigure}
\caption{(a) A noisy circle $X$ mapped to the $x$-axis and covered by three explicit intervals. (b) Clustering within the preimages yields a simplicial complex that preserves the 1-dimensional hole. (c) Merging without clustering collapses the graph structure to a tree.}
\label{fig:mapper-circle-example}
\end{figure}
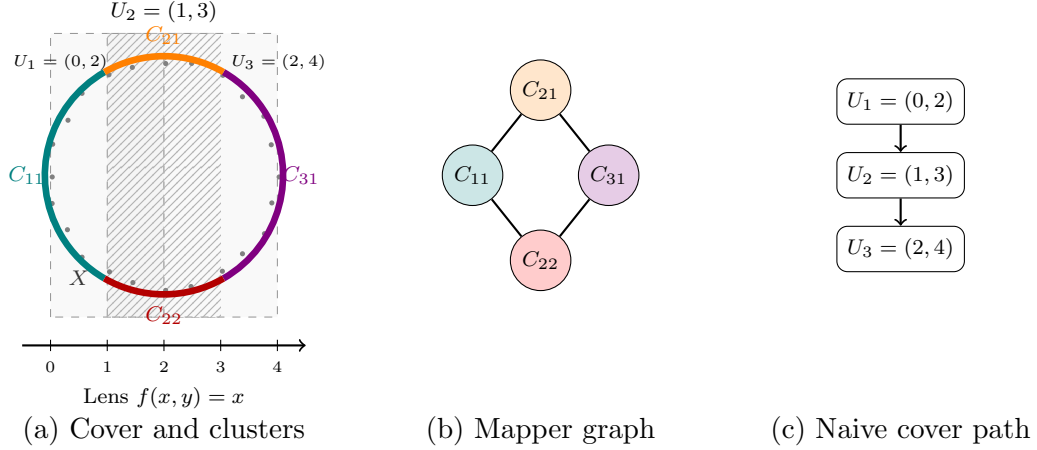

\textbf{Example.} Let $X$ be a noisy sample of a circle in $\mathbb{R}^2$, with lens $f(x,y) = x$. Cover $f(X)$ with three overlapping intervals $U_1=(0,2)$, $U_2=(1,3)$, and $U_3=(2,4)$ (Figure~\ref{fig:mapper-circle-example}\subref{fig:cover-clusters}). Since $U_1$ and $U_3$ each contain an extremal value of $f$, their preimages are single connected arcs, giving one cluster each: $f^{-1}(U_1)=C_{11}$ and $f^{-1}(U_3)=C_{31}$. The middle interval $U_2$ excludes both extremes, so $f^{-1}(U_2)$ splits into a top and a bottom arc, yielding two clusters $C_{21}$ and $C_{22}$. Overlaps between the intervals mean $C_{11}$ and $C_{31}$ each connect to both $C_{21}$ and $C_{22}$, producing a 4-cycle that recovers the loop (Figure~\ref{fig:mapper-circle-example}\subref{fig:mapper-graph}). Skipping clustering and using the cover elements as vertices instead collapses the graph to the path $U_1\text{--}U_2\text{--}U_3$, losing the loop (Figure~\ref{fig:mapper-circle-example}\subref{fig:naive-path}).

\subsubsection{Clustering within Cover Elements}

For each cover element $U_i$, Mapper restricts attention to the preimage 
$f^{-1}(U_i) = \{x \in X : f(x) \in U_i\}$ and applies a clustering algorithm locally to identify dense subsets of observations. Mapper often uses Density-Based Spatial Clustering of Applications with Noise (DBSCAN)~\cite{ester1996density}.
The algorithm is governed by two parameters: $\varepsilon_{\text{DBSCAN}} > 0$, controlling 
the neighborhood radius, and \textit{min\_samples} $\in \mathbb{Z}^+$, 
the minimum number of points required to form a dense region. For a point 
$Z_i$, its $\varepsilon_{\text{DBSCAN}}$-neighborhood is defined as
$N_{\varepsilon_{\text{DBSCAN}}}(Z_i) = \{Z_j : d(Z_i, Z_j) < \varepsilon_{\text{DBSCAN}}\},$ where $d$ denotes a chosen distance metric.

\noindent\textbf{Mapper as a Dimensionality Reduction Tool.}

\begin{figure}[H]
    \centering
    \begin{subfigure}[b]{0.40\textwidth}
        \centering
        \includegraphics[width=\textwidth]{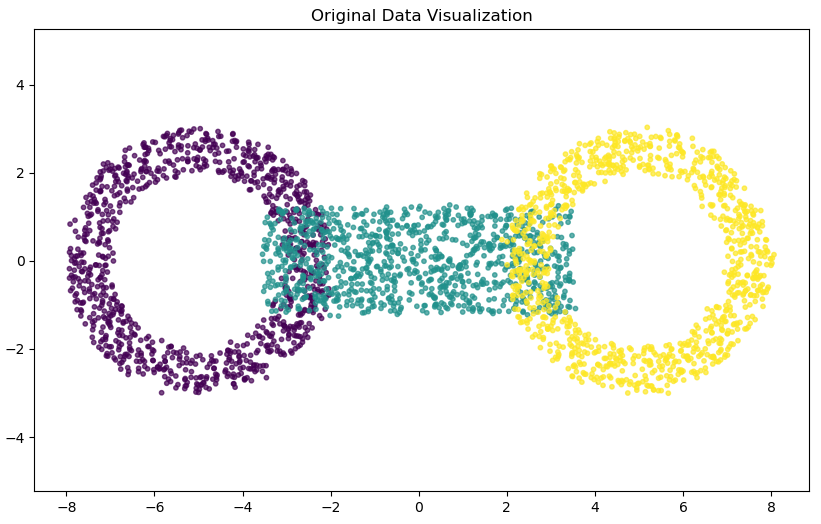}
        \caption{Original point cloud.}
        \label{fig:Mapper-raw}
    \end{subfigure}
    \hfill
    \begin{subfigure}[b]{0.35\textwidth}
        \centering
        \includegraphics[width=\textwidth]{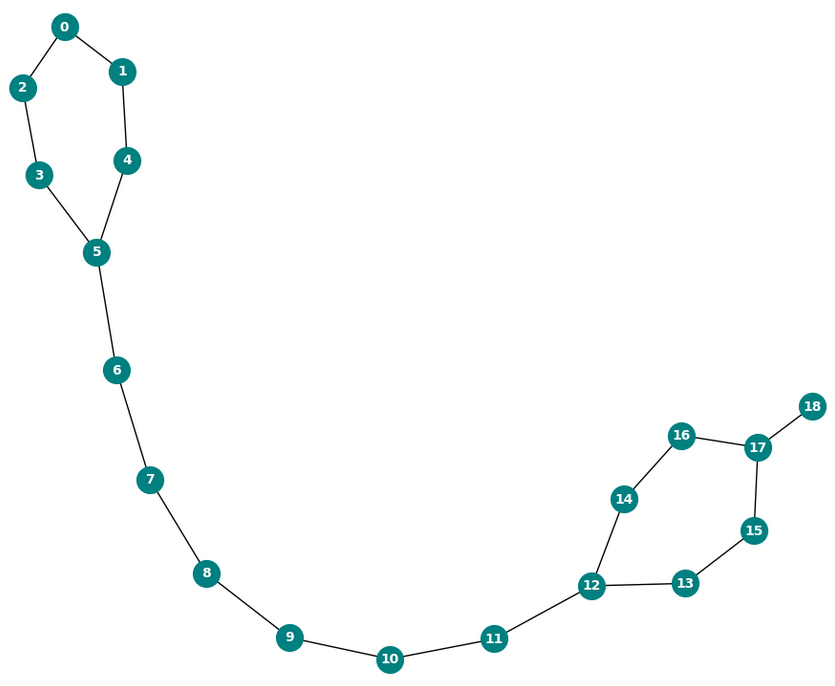}
        \caption{Recovered Mapper graph.}
        \label{fig:Mapper-graph}
    \end{subfigure}
    \caption{Mapper applied to the noisy annuli--bridge dataset.}
    \label{fig:Mapper-side-by-side}
\end{figure}

Figure~\ref{fig:Mapper-side-by-side} illustrates the dimensionality reduction 
capacity of the Mapper algorithm where the 3,000-point noisy annuli--bridge 
dataset in Figure~\ref{fig:Mapper-side-by-side}\subref{fig:Mapper-raw} is compressed into a compact graph of just 19 vertices and 
20 edges in Figure~\ref{fig:Mapper-side-by-side}\subref{fig:Mapper-graph}, yet the two-loops-plus-bridge topology of the original point 
cloud is faithfully preserved, demonstrating that Mapper can drastically 
reduce data complexity without sacrificing the essential geometric and 
topological structure of the data.

\subsection{Survival Analysis}

Survival analysis is the statistical analysis that examines time-to-event outcomes in which the event of interest may not be observed for every subject by the time the study ends, a phenomenon known as \textit{censoring}. Let $T \geq 0$ denote the survival time and $\delta \in \{0,1\}$ the event indicator, where $\delta = 1$ denotes an observed event and $\delta = 0$ a censored observation. Three complementary methods are employed.

\subsubsection{Cox Proportional Hazards Model}
Let $t_1 < t_2 < \ldots < t_k$ denote the event times, where $d_j$ number of events occur at the time $t_j$, and there are $n_j$ subjects still surviving (at risk) just before the time $t_j$.
The {Kaplan--Meier} estimator~\cite{kaplan1958nonparametric} of the survival function, which provides the survival probability $S(t)=P(T>t)$, is given by: 
$$
    \widehat{S}(t)
    = \prod_{j:\, t_{(j)} \leq t}
      \left(1 - \frac{d_j}{n_j}\right).
$$

\begin{figure}[H]
    \centering
    \begin{subfigure}[b]{0.48\textwidth}
        \centering
        \begin{tikzpicture}[scale=0.75]
        \begin{axis}[
            xlabel={Time},
            ylabel={$\widehat{S}(t)$},
            xmin=0, xmax=14,
            ymin=0, ymax=1.1,
            ytick={0, 0.2, 0.4, 0.6, 0.8, 1.0},
            grid=major,
            grid style={dashed, gray!30},
        ]
        \addplot[const plot, thick, blue] coordinates {
            (0,1.0)(3,0.9)(4,0.7)(6,0.6)
            (7,0.4)(8,0.3)(10,0.2)(11,0.1)
            (13,0.0)(14,0.0)
        };
        \end{axis}
        \end{tikzpicture}
        \caption{Single-group KM curve.}
        \label{fig:km_single}
    \end{subfigure}
    \hfill
    \begin{subfigure}[b]{0.48\textwidth}
        \centering
        \begin{tikzpicture}[scale=0.75]
        \begin{axis}[
            xlabel={Time},
            ylabel={$\widehat{S}(t)$},
            xmin=0, xmax=16,
            ymin=0, ymax=1.1,
            ytick={0, 0.2, 0.4, 0.6, 0.8, 1.0},
            grid=major,
            grid style={dashed, gray!30},
            legend pos=north east,
        ]
        \addplot[const plot, thick, blue] coordinates {
            (0,1.0)(1,0.9)(2,0.7)(3,0.5)
            (4,0.3)(5,0.2)(6,0.1)(7,0.0)
        };
        \addlegendentry{Group 1}
        \addplot[const plot, thick, red] coordinates {
            (0,1.0)(5,0.9)(6,0.8)(8,0.7)
            (9,0.6)(11,0.4)(12,0.3)(13,0.3)
        };
        \addlegendentry{Group 2}
        \end{axis}
        \end{tikzpicture}
        \caption{Two-group KM curves ($p < 0.05$).}
        \label{fig:km_twogroup}
    \end{subfigure}
    \caption{Kaplan--Meier survival curves.}
    \label{fig:km_curves}
\end{figure}
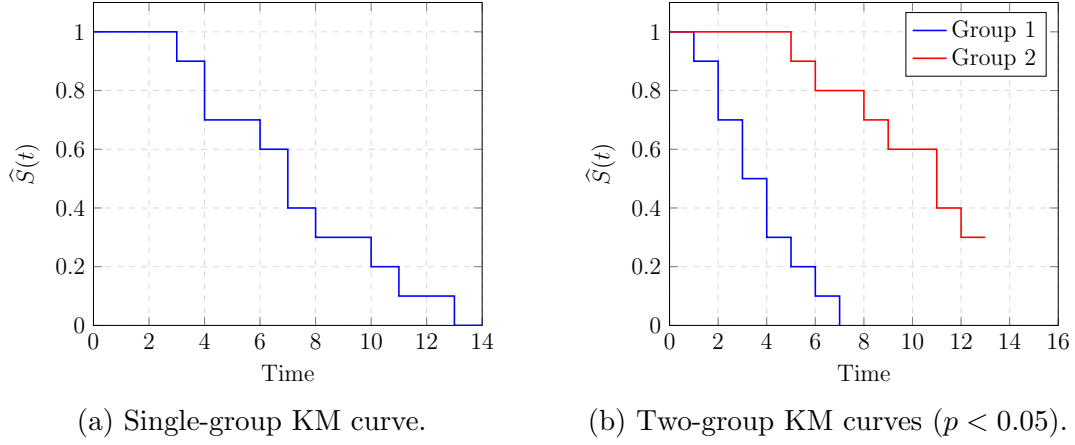

The resulting estimator is visualised as a step function called the {Kaplan--Meier curve} which decreases each time an event occurs and remains flat between events, providing a non-parametric estimate of the survival distribution over time as shown in Figure~\ref{fig:km_curves}\subref{fig:km_single}.

\subsubsection{Log-Rank Test}
The log-rank test is used to check for statistical group-level 
differences in survival. At each distinct event time $t_{(j)}$, 
the observed and expected events in group 1, $d_{1j}$ and $e_{1j}$, 
are compared via the log-rank statistic
$$
\chi^2_{\mathrm{LR}} = \frac{\left(\sum_j(d_{1j} - e_{1j})\right)^2}
{\sum_j V_j} = \frac{\left(\sum_j(d_{2j} - e_{2j})\right)^2}
{\sum_j V_j},
$$
where $V_j$ is the hypergeometric variance,
$$
V_j = \frac{n_{1j} n_{2j} d_j (n_j - d_j)}{n_j^2 (n_j - 1)},
$$

with $n_{1j}$, $n_{2j}$ the number at risk in groups 1 and 2, 
$n_j = n_{1j} + n_{2j}$, and $d_j$ the total events at $t_{(j)}$. 
The algebraic constraint $d_{1j} - e_{1j} = -(d_{2j} - e_{2j})$ 
guarantees that both formulations yield an identical statistic, 
confirming that the deviation in either group fully captures the 
discrepancy between them, while $n_{2j}$ in $V_j$ reflects group 2's 
direct contribution to the variance. Under $H_0: S_1(t) = S_2(t)$, 
$\chi^2_{\mathrm{LR}}$ follows a $\chi^2$ distribution with one degree 
of freedom, and the $p$-value is obtained as 
$p = P(\chi^2_1 > \chi^2_{\mathrm{LR}})$, the upper-tail probability 
of the $\chi^2_1$ distribution; $p < 0.05$ indicates a statistically 
significant difference between the survival distributions, leading to 
rejection of $H_0$. 

When comparing more than two groups, the log-rank test is generalized using a vector of deviations and a variance--covariance matrix. The deviation vector is

$$
\mathbf{U} = \mathbf{O} - \mathbf{E} =
\begin{bmatrix}
O_1 - E_1 \\
O_2 - E_2 \\
\vdots \\
O_k - E_k
\end{bmatrix},
$$
where $O_g = \sum_j d_{gj}$ and $E_g = \sum_j e_{gj}$ are the total 
observed and expected events in group $g$, indicating which groups 
had more or fewer events than expected. The variance--covariance 
matrix
$$
\mathbf{V} =
\begin{bmatrix}
\mathrm{Var}_1 & \mathrm{Cov}_{12} & \cdots & \mathrm{Cov}_{1k} \\
\mathrm{Cov}_{21} & \mathrm{Var}_2 & \cdots & \mathrm{Cov}_{2k} \\
\vdots & \vdots & \ddots & \vdots \\
\mathrm{Cov}_{k1} & \mathrm{Cov}_{k2} & \cdots & \mathrm{Var}_k
\end{bmatrix}
$$
captures how deviations across groups are linked, since all groups 
share the same risk set at each event time. Because $\sum_{g=1}^{k}(O_g - E_g) = 0$ at every event time, the rows and columns of $\mathbf{V}$ are linearly dependent and $\mathbf{V}$ is singular; in practice one group is dropped, reducing $\mathbf{U}$ and $\mathbf{V}$ to dimension $k-1$ before inversion (equivalently, a generalized inverse $\mathbf{V}^{-}$ may be used on the full $k \times k$ system). These quantities combine 
into the test statistic
$$
\chi^2_{\mathrm{LR}} = \mathbf{U}^T \mathbf{V}^{-1} \mathbf{U},
$$
which, under $H_0: S_1(t) = S_2(t) = \dots = S_k(t)$, follows a 
$\chi^2$ distribution with $k-1$ degrees of freedom,
$$
\chi^2_{\mathrm{LR}} \sim \chi^2_{k-1}.
$$
The $p$-value is obtained as $p = P(\chi^2_{k-1} > \chi^2_{\mathrm{LR}})$, 
the upper-tail probability of the $\chi^2_{k-1}$ distribution; 
$p < 0.05$ indicates a statistically significant difference among the 
survival distributions, leading to rejection of $H_0$.

\subsubsection{Cox Proportional Hazards Model}
The Cox proportional hazards model is used to analyze the impact of multiple predictors on survival outcomes by estimating their effect on the hazard rate. The predictor variables are generally referred to as covariates in survival analysis.
To adjust for confounding covariates, the Cox model~\cite{cox1972regression}
specifies
\begin{equation*}
  h(t \mid \mathbf{x}) \;=\; h_{0}(t)\exp\!\left(\boldsymbol{\beta}^{\!\top}
  \mathbf{x}\right),
\end{equation*}
where $h_{0}(t)$ is an unspecified baseline hazard and
$\boldsymbol{\beta}$ is estimated by maximising the partial log-likelihood $\ell(\beta)$ as shown below;

$$U(\beta) = \frac{\partial  \ell (\beta)}{\partial \beta} = 0; \qquad \ell(\beta) = \displaystyle \sum_{k = 1}^D \left[ \beta x_{(k)} - 
\ln \displaystyle \sum_{j \in R(t_k)} e^ {\beta x_{j}} \right],$$

where $\ell(\beta)$ is the partial log-likelihood, $\beta$ is the coefficient, $D$ is the total events, $x_{(k)}$ is the covariate value of the individual experiencing the event at time $t_k$, $x_j$ indicates if individual $j$ belongs to the comparison group, and $R(t_k)$ is the risk set.

\subsubsection{Hazard Ratio Interpretation.}
The hazard ratio for covariate $k$ is defined as the ratio of hazards 
between two subjects differing by one unit in $x_k$, with all other 
covariates held fixed. From the Cox model,
$$
\mathrm{HR}_k = \frac{h(t \mid x_k+1)}{h(t \mid x_k)}
= \frac{h_0(t)\exp\!\big(\beta_k(x_k+1) + \displaystyle\sum_{l \neq k}\beta_l x_l\big)}
       {h_0(t)\exp\!\big(\beta_k x_k + \displaystyle\sum_{l \neq k}\beta_l x_l\big)}
= \exp(\beta_k),
$$
so that the baseline hazard $h_0(t)$ and all other covariates cancel, leaving the hazard ratio dependent only on $\beta_k$. Substituting the maximum partial-likelihood estimate $\hat\beta_k$ gives $\widehat{\mathrm{HR}}_k = \exp(\hat\beta_k)$, which quantifies the multiplicative change in instantaneous risk per unit increase in covariate $k$: $\mathrm{HR}_k > 1$ indicates elevated risk, $\mathrm{HR}_k < 1$ a protective effect, and $\mathrm{HR}_k = 1$ implies no effect.

Statistical significance of $\hat\beta_k$ is assessed via the Wald test statistic
$$
Z_k = \frac{\hat\beta_k}{\mathrm{SE}(\hat\beta_k)},
$$
where $\mathrm{SE}(\hat\beta_k)$ is the standard error obtained from the square root of the corresponding diagonal entry of the inverse observed information matrix, $$\mathrm{SE}(\hat\beta_k) = \sqrt{\big[I(\hat\beta)^{-1}\big]_{kk}},$$ with $I(\beta) = -\frac{\partial^2\ell(\beta)}{\partial\beta\partial\beta^\top}$. Under $H_0: \beta_k = 0$, $Z_k$ follows a standard normal distribution asymptotically, and the two-sided $p$-value is $p = 2\big(1-\Phi(|Z_k|)\big)$, where $\Phi$ is the standard normal cumulative distribution function. A confidence interval for the hazard ratio follows directly as $\exp\big(\hat\beta_k \pm 1.96\,\mathrm{SE}(\hat\beta_k)\big)$. A value of $p < 0.05$ indicates that covariate $k$ has a significant independent effect on survival after adjusting for all other covariates.

\section{Results and Discussions}
\subsection{Mapper Network Reproduction}

As a first step, the Mapper framework of Rostami et al.~\cite{rostami2025tda} was reconstructed on the TCGA-BRCA dataset, comprising both a sample network, in which nodes represent clusters of patients with similar gene expression profiles, and a feature network, in which nodes represent clusters of co-expressed genes. Reproducing both networks confirms that the topological structures reported in the original study are recoverable within this pipeline, providing a validated foundation for the survival analysis presented in subsequent subsections.

\subsubsection{Topological sample network of the subtypes}
To validate pipeline reproducibility, the topological sample network of Rostami and colleagues \cite{rostami2025tda} was reconstructed by applying the Mapper algorithm to TCGA-BRCA gene expression data, using a two-dimensional lens combining $L_1$ (Manhattan) eccentricity and the first principal component of the expression matrix, a cover of 12 intervals with 50\%
overlap, and DBSCAN clustering (Euclidean distance, $\varepsilon = 55$, min\_samples $= 2$), with each node assigned its dominant clinical subtype label whenever that subtype accounted for over 65\% of its members, and labeled ``mixed" otherwise. Three structural patterns observed in the original study were recovered in this network: the Basal-like subtype forms an isolated, well-separated branch; Luminal A occupies a dense hub interspersed with Normal-like nodes; and HER2-enriched nodes are positioned between these two regions, consistent with its intermediate molecular profile. The recovery of these patterns supports the reproducibility of the original topological structure within this pipeline.

\begin{figure}[H]
    \centering
    \includegraphics[scale=0.6]{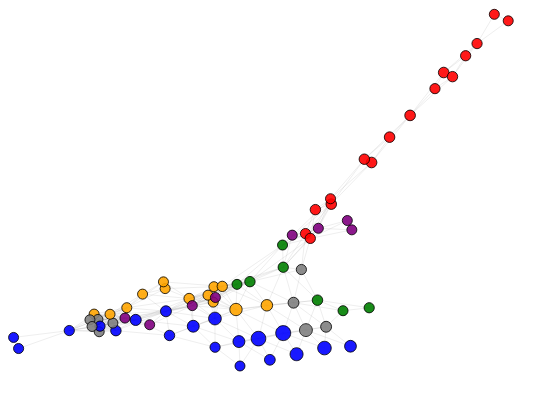}\\[1pt]
    \includegraphics[scale=0.6]{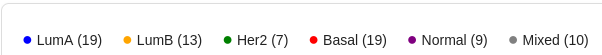}
    \caption{Reproduced TDA-based breast cancer patient sample network matching the topology of Rostami and colleagues~\cite{rostami2025tda}, where node sizes reflect cluster membership and colors denote dominant clinical subtypes.}
    \label{fig:sample_Mapper}
\end{figure}

\subsubsection{Feature Network}

\begin{figure}[H]
  \centering
  \begin{subfigure}[t]{0.33\linewidth}
    \includegraphics[width=\linewidth]{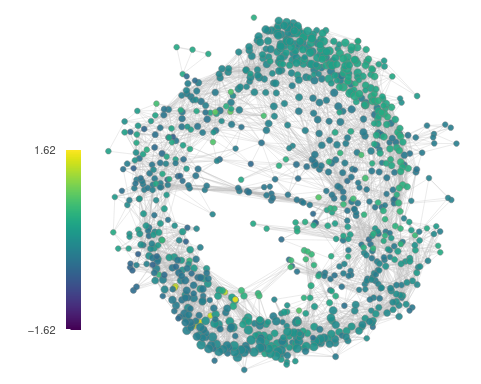}
    \caption{}
    \label{fig:HER2}
  \end{subfigure}\hfill
  \begin{subfigure}[t]{0.33\linewidth}
    \includegraphics[width=\linewidth]{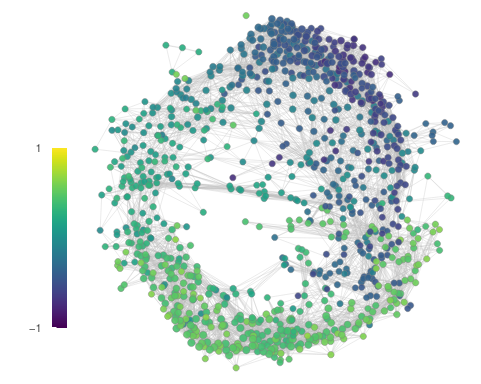}
    \caption{}
    \label{fig:luma}
  \end{subfigure}\hfill
  \begin{subfigure}[t]{0.33\linewidth}
    \includegraphics[width=\linewidth]{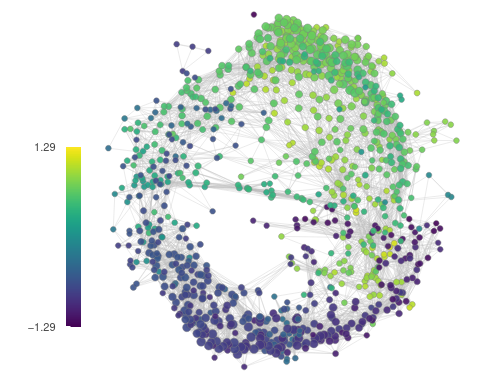}
    \caption{}
    \label{fig:basal}
  \end{subfigure}
  \vspace{-2pt}
  \caption{Feature network where nodes represents clusters of highly correlated genes, colored by breast cancer subtypes to reveal distinct co-expression regions. (a) In HER2-overexpressing (IHC 3+) patients, highly correlated genes in the lower network segment are strongly expressed (yellow), while the upper segment is suppressed (purple). (b) Luminal A patients display a similar trend, with high gene activation concentrated among correlated genes in the lower network segment. (c) Basal-like patients show an inverted pattern, where highly correlated genes in the upper segment are strongly activated and those in the lower segment are suppressed.}
 \label{fig:feature_networks}
\end{figure}

To construct the gene feature network, the gene expression matrix was transposed so that genes become the data points, and Mapper was reapplied using correlation distance, grouping genes into the same node only when their expression profiles were highly correlated across patients. Each node was colored by the mean $z$-score of a given patient subgroup to identify which gene modules are most active in that subtype (HER2 IHC 3+ here denotes clinical HER2 positivity by immunohistochemistry, distinct from the PAM50 HER2-enriched molecular subtype). Figure~\ref{fig:feature_networks} shows the same gene co-expression network colored by three breast cancer subtypes. HER2 IHC 3+ (Figure~\ref{fig:HER2}) and Luminal A (Figure~\ref{fig:luma}) patients both show strong activation (yellow) concentrated in the lower network segment, with the upper segment suppressed (purple), indicating shared gene activity between these subtypes. Coloring by Basal-like patients (Figure~\ref{fig:basal}) reverses this pattern entirely: the upper segment is strongly activated while the lower segment is suppressed. This reversal demonstrates that Basal-like gene modules, linked to high proliferation, DNA damage response, and loss of Luminal differentiation, are topologically separated from the Luminal A- and HER2-associated modules, confirming a clear separation of subtype-specific co-expression programs within the feature network.

\subsubsection{Low-Dimensional visualization}
\begin{figure}[H]
  \centering
  \includegraphics[scale=0.26]{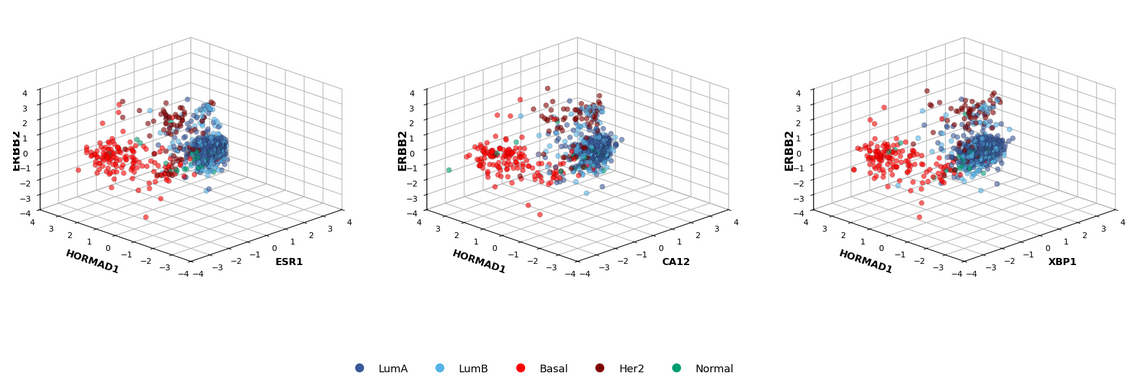}
  \vspace{-6pt}
  \caption{Low-dimensional visualization of TCGA BRCA patients. 3D scatter plots colored by PAM50 subtype across \textit{ERBB2} (vertical), \textit{HORMAD1} (left horizontal), and a varying right axis (\textit{ESR1}, \textit{CA12}, \textit{XBP1}). Basal-like patients (red) isolate along \textit{HORMAD1}, HER2-enriched (dark red) elevate along \textit{ERBB2}, and Luminal subtypes (blue) cluster near the origin.}
  \label{fig:scatter3d}
\end{figure}

Using genes identified directly from the feature network, 3D scatter plots were constructed for all patients: \textit{ERBB2} (vertical axis) from the HER2-enriched peripheral cluster, \textit{HORMAD1} (shared left horizontal axis) from the brightest nodes of the Basal-like upper arc, and a rotating right axis (\textit{ESR1}, \textit{CA12}, \textit{XBP1}) from the Luminal~A lower arc. In all three panels of Figure~\ref{fig:scatter3d}, Basal-like patients (red) consistently separate from the rest of the cohort along \textit{HORMAD1} regardless of which Luminal~A gene occupies the third axis, confirming it as a robust topology-derived separator, while HER2-enriched patients (dark red) are elevated along \textit{ERBB2} and Luminal~A/B (dark/light blue) cluster near the origin. Notably, these four topology-derived genes alone achieve a subtype separation that conventional classifiers require the full 50-gene PAM50 panel to approximate.

\subsection{Topological Position and Patient Survival}
This subsection examines whether a patient's position within the Mapper graph measured through distance to high-risk landmarks, node-level proliferation activity, and localisation at subtype boundaries carries information about survival beyond conventional PAM50 classification.

\subsubsection{Node Distance to High-Risk Landmark Node}

\begin{figure}[H]
    \centering
    \includegraphics[scale=0.50]{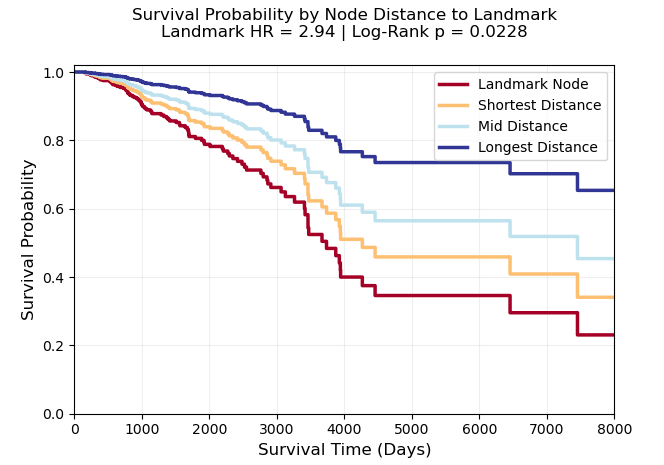}
    \caption{KM curves for the Node Distance to High-Risk Landmark Node}
    \label{fig:node_distance}
\end{figure}

A landmark node was identified as the node with the highest hazard ratio (HR = 2.94) obtained from a series of univariate Cox proportional-hazards models, each fit with a single node's membership (patients within that node vs.\ all other patients) as the sole predictor of survival; this hazard ratio therefore represents the relative event rate for patients within a given node compared to all patients outside it. Node distance to this landmark was then grouped by tercile into shorter, mid, and longer distance categories. Figure~\ref{fig:node_distance} shows that survival worsens with proximity to the landmark ($p = 0.0228$, log-rank test on tercile groups), but this signal
did not survive adjustment for age and stage: the continuous-distance HR moved from 0.90 ($p = 0.075$) to 1.01 ($p = 0.90$) once both covariates were added, with age and stage remaining the dominant predictors ($p < 0.0001$ and $p < 5\times10^{-5}$). However, landmark distance should not be interpreted as an independent prognostic marker, since its effect disappeared after adjustment for age and stage. This is explained by demographics across distance groups: mean age reduced from 71.3 to 54.4 years moving from the landmark node to the longest distance, and modal stage shifted from IIIA at the landmark to IIA. The result's value is not landmark distance as an independent marker, but as evidence that Mapper, built from gene expression alone with no survival input, recovers a risk gradient that tracks real clinical outcome, useful as a proxy stratifier where age or stage data are unavailable.

\subsubsection{Effect of Node Proliferation Score on Patient Survival}

\begin{figure}[H]
    \centering
    \includegraphics[scale = 0.20]{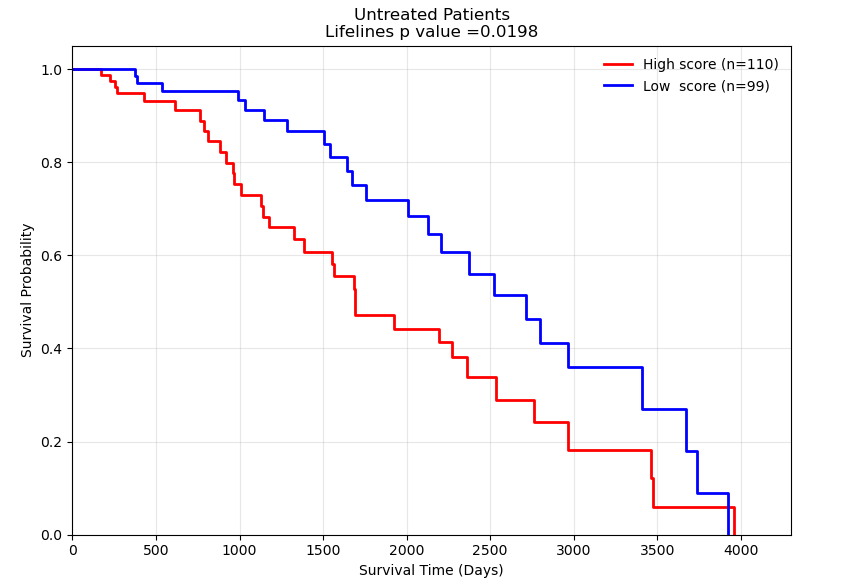}
    \caption{KM curves of untreated patients from high‑proliferation and low‑proliferation nodes.}
    \label{fig:untreated}
\end{figure}

For each Mapper node, the node's proliferation score was defined as the mean of the six proliferation-gene patient-level scores (\textit{MKI67}, \textit{CCNB1}, \textit{MYBL2}, \textit{BIRC5}, \textit{UBE2C}, \textit{RRM2}) among patients whose average node-gene expression exceeded the node's median, these six genes being established markers of cell-cycle activity and tumor proliferation~\cite{liu2018identification}. Nodes scoring at or above the median across all nodes were classified as high proliferation and the rest as low; patients were then assigned to the high-proliferation group via a voting mechanism if they appeared in at least as many high- as low-proliferation nodes, and to the low-proliferation group otherwise. Clinically, this shows that purely topological structure derived from gene expression alone can recover a proliferation axis with established prognostic and treatment-response relevance, without proliferation markers informing the graph's construction. To isolate this signal from treatment effects, the analysis was first restricted to untreated patients, since therapy alters both gene expression and survival trajectories. The Kaplan--Meier curves in Figure~\ref{fig:untreated} show that the high-proliferation group (red) had lower survival than the low-proliferation group (blue), confirming that higher proliferation-gene expression is linked to worse outcomes.

\begin{figure}[H]
    \centering
    \includegraphics[scale = 0.3]{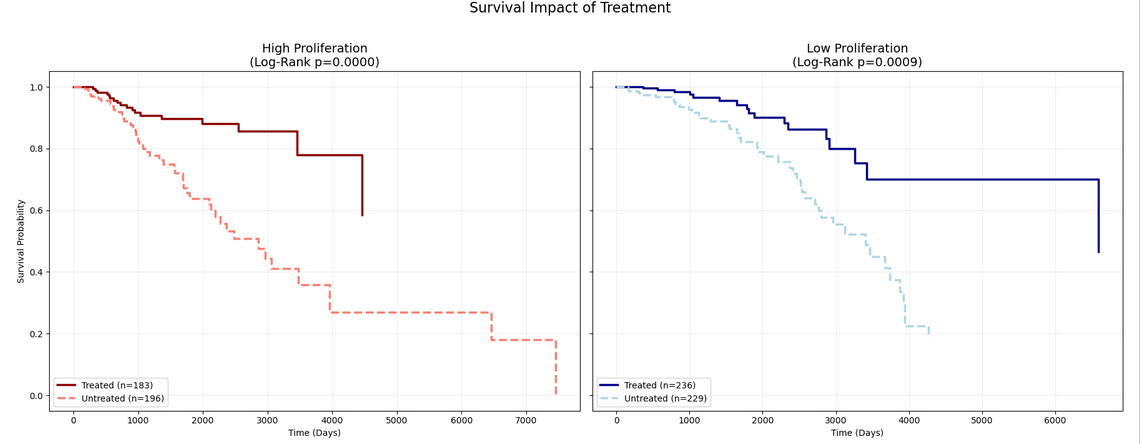}
    \caption{The effect of treatment on Survival}
    \label{fig:treated}
\end{figure}

\noindent\textbf{Effect of Treatment Across Proliferation Groups:} The 
Kaplan--Meier curves in Figure~\ref{fig:untreated} show that among untreated patients, the high proliferation group had lower survival than the low proliferation group. Among treated patients (Figure~\ref{fig:treated}), both high and low proliferation patients showed improved survival, with treated groups doing significantly better than untreated ones (log-rank $p < 0.0001$), indicating that treatment narrowed the survival gap between the two proliferation groups.

\subsubsection{Identification of Discordant Survivors and Their Gene Signatures via Mapper Topology}

\begin{figure}[H]
    \centering
    \includegraphics[scale=0.25]{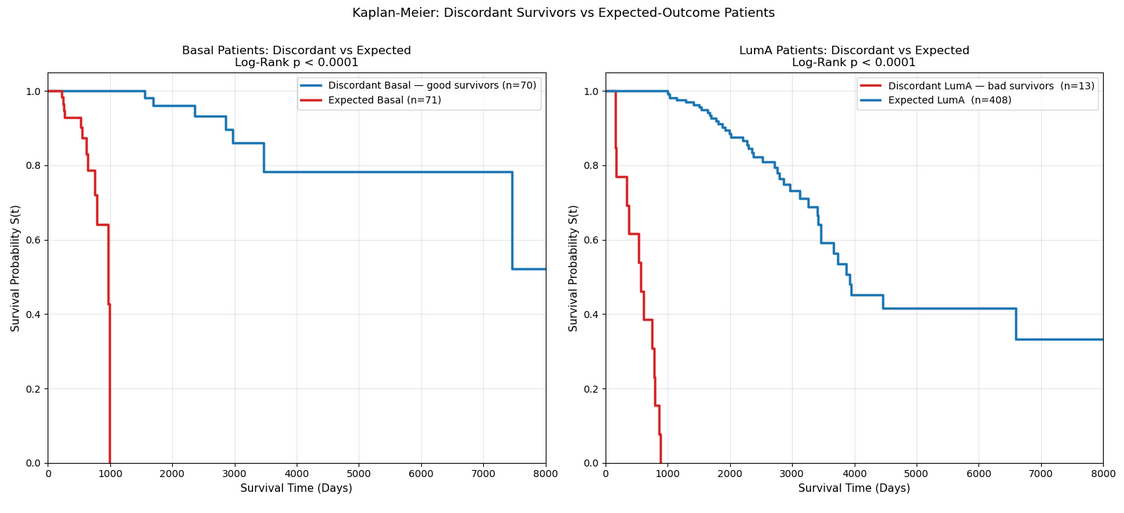}
    \caption{KM curves of the Discordant survivors and the expected outcome patients.}
    \label{fig:discondant}
\end{figure}

Mapper topology can identify survival discordance beyond traditional PAM50 classification. The Discordant Basal group includes Basal-like patients who survived longer than the median Luminal A patient (unexpectedly good survival for a high-risk subtype), while the Discordant Luminal A group includes Luminal A patients who died before the median Basal-like survival time (unexpectedly poor survival for a low-risk subtype). The Kaplan--Meier curves in Figure~\ref{fig:discondant} show clear survival separation, with Discordant Basal patients surviving much longer than expected Basal patients and Discordant Luminal A patients surviving much shorter than expected Luminal A patients.

\begin{figure}[H]
    \centering
    \includegraphics[scale=0.45]{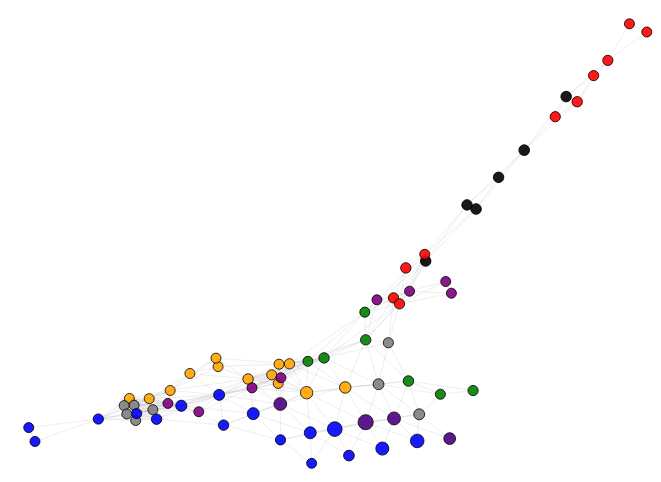}
    \caption{Mapper graph highlighting the nodes most enriched for discordant survivors. Black nodes are Basal-majority nodes dominated by discordant Basal survivors; deep purple nodes are enriched for discordant Luminal A patients; grey nodes show little or no enrichment for either group.}
    \label{fig:discordant_hotspots}
\end{figure}

Discordant patients were located near transitional and mixed nodes at subtype boundaries rather than in subtype cores, as shown in Figure~\ref{fig:discordant_hotspots}. Cox models showed that discordant assignment remained highly significant even after adjusting for age and pathological stage: in Basal-like patients, the Discordant Flag predicted improved survival (HR = 0.223, $p$ = 0.0005), with stage also significant (HR = 1.835, $p$ = 0.0259) and age non-significant, while in Luminal A patients, the Discordant Flag predicted poor survival (HR = 65.91, $p < 0.0001$), with stage and age non-significant.

\begin{figure}[H]
  \centering
  \begin{subfigure}[t]{0.35\linewidth}
    \includegraphics[scale=0.25]{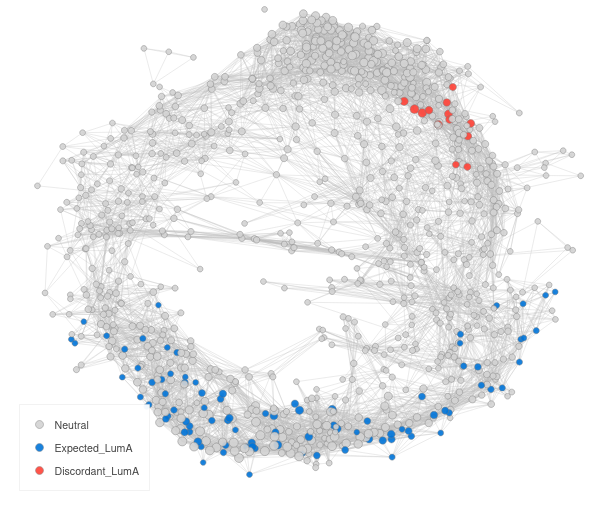}
    \caption{}
  \end{subfigure}\hfill
  \begin{subfigure}[t]{0.35\linewidth}
    \includegraphics[scale = 0.25]{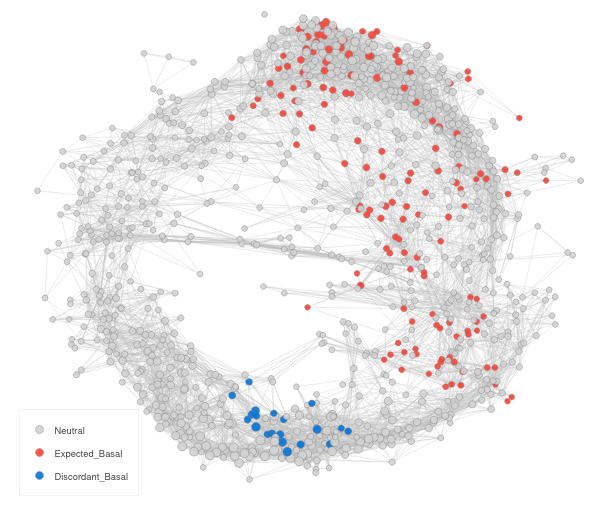}
    \caption{}
  \end{subfigure}\hfill
  \vspace{-2pt}
  \caption{Visualization of gene networks with discordant expression among breast cancer phenotypes. (a) Discordant Luminal A (red) genes form a distinct cluster in the upper part of the feature Mapper, showing expression differences from Expected Luminal A (blue). (b) Discordant Basal (blue) genes form a distinct cluster in the lower part, while Expected Basal (red) genes are located in the upper right corner.}
 \label{fig:discordant_genes}
\end{figure}

To interpret the biology behind this, treated Basal and Luminal A patients were divided into Expected and Discordant groups, and differentially expressed genes were identified using $Log2$ Fold Change ($Log2FC$)~\cite{love2014moderated}, with a small constant $\epsilon = 0.00001$ added to avoid division by zero. As shown in Figure~\ref{fig:discordant_genes}, the top differentially expressed genes revealed a cross-subtype signature reversal: Discordant Luminal A genes (\textit{PDK1}~\cite{xie2022pdk1}, \textit{EIF2C2}~\cite{casey2019ago2}, \textit{CSNK2A2}~\cite{bae2016csnk2a1}, \textit{TPI1}~\cite{jin2022tpi1}, \textit{NUP93}~\cite{nataraj2022nup93,bersini2020nup93}, \textit{XPOT}, \textit{C16orf80}, \textit{SF3B3}, \textit{SLC25A13}, and \textit{GARS}) clustered in the upper, Basal-associated segment of the feature network, several with documented roles in glycolysis, invasion, and reduced survival, suggesting an aberrant aggressive gene program, while Discordant Basal genes (\textit{AKAP12}~\cite{soh2018akap12}, \textit{TCEAL7}~\cite{chien2008tceal7}, \textit{MYH11}~\cite{cerqueira2022myh11}, \textit{KANK2}, \textit{SMOC2}, \textit{ZCCHC24}, \textit{ZNF423}, \textit{ITGB5}, \textit{MAB21L1}, and \textit{LCA5}) localized to the lower, Luminal A-associated segment, several known as tumor suppressors or markers of a differentiated state, reflecting a more differentiated state that may explain their unexpectedly favorable survival. This suggests Mapper captures candidate treatment-response-related biology not captured by PAM50 classification alone; the Basal-like program in Discordant Luminal A patients may explain hormone-therapy resistance, while the Luminal A-like program in Discordant Basal patients may explain their greater treatment sensitivity.

\section{Train-Test Validation for Topological Risk Stratification}

We next test whether Mapper topology generalizes beyond the patients used to build it, by constructing the Mapper graph from 80\% of TCGA BRCA patients and evaluating risk stratification on the held-out 20\%. Clinical variables were withheld until after risk groups were assigned, preventing data leakage. The event variable was set to 1 with survival time as days to death if death occurred, and 0 with survival time as last follow-up otherwise.

\subsection{Construction of the Training Mapper Graph and Risk Tier Derivation}

The Mapper graph was built using only the 80\% training group, with a two-dimensional lens combining $L_1$ eccentricity (average Manhattan distance from each patient to the rest of the cohort) and the first principal component of the training gene expression matrix. The cover consisted of 12 intervals with 50\% overlap, and clustering was performed with DBSCAN using Euclidean distance, $\varepsilon = 55$, and min\_samples $= 5$. A univariate Cox model was fitted for each node using a binary in-node indicator, with the hazard ratio (HR) indicating whether patients in that node died faster (HR $>$ 1) or slower (HR $<$ 1) than the rest. Node HRs were rank-transformed and split into tertiles: High Risk (upper), Mid Risk (middle, including HR values near 1.0), and Low Risk (lower). Each training patient was assigned the risk tier of
their highest-HR node among all nodes they belong to; when a patient's nodes tied for the highest HR, the tier of the first such node encountered was retained. For transfer to validation patients, each validation patient's gene expression profile was correlated (Pearson) against the mean expression profile of every training node, and matched to the node with the highest
positive correlation; if this correlation exceeded 0.65, the patient inherited that node's risk tier directly, otherwise the tier was determined by a Borda-count ranking across all positively correlated nodes.

\subsubsection{Kaplan--Meier Survival Analysis}

\begin{figure}[H]
    \centering
    \includegraphics[scale=0.35]{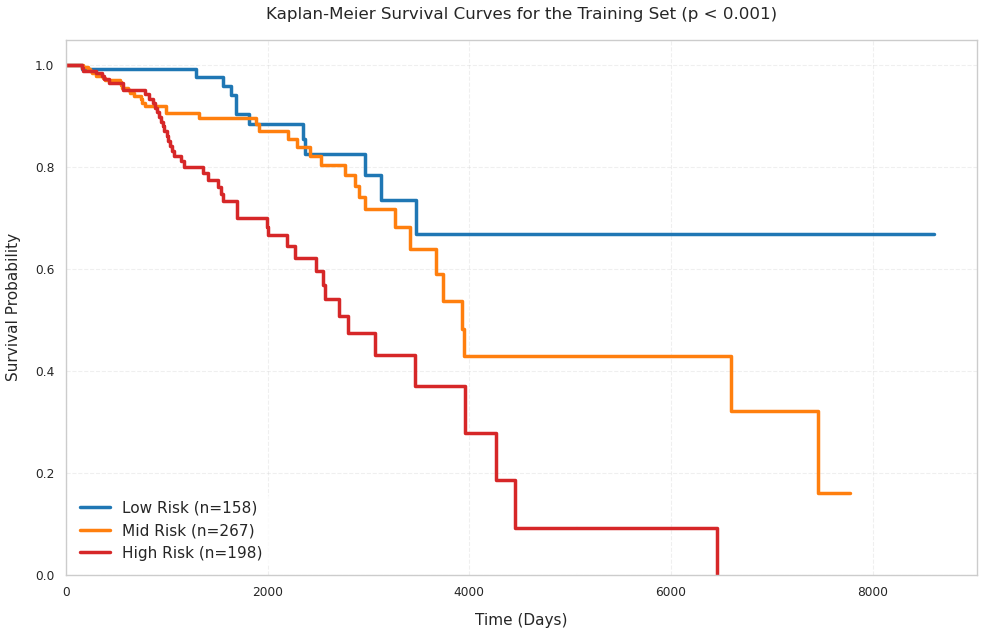}
    \caption{KM curves for the training group}
    \label{fig:KM_train}
\end{figure}

Survival analysis using Kaplan--Meier curves for the three Mapper risk tiers (Figure~\ref{fig:KM_train}) showed clear monotonic separation from early follow-up times onward, maintained throughout the entire observation window. High Risk patients (red) had the fastest survival decline, linked to aggressive, hyperproliferative tumors. Low Risk patients (blue) showed the slowest decline, reflecting stable low-proliferation biology. Mid Risk patients had an intermediate curve, decreasing faster than Low Risk but slower than High Risk. The multivariate log-rank test was statistically significant ($p < 0.001$), validating the strength of the topological risk categories in predicting patient survival.

\subsubsection{Multivariable Cox modeling of the Training Set}
Two multivariable Cox models tested whether the Mapper risk tiers held prognostic value independent of standard clinical variables. Stage was significant across all three tiers, while age, treatment, and proliferation score mattered mainly within the High Risk tier, with treatment significantly reducing hazard there, suggesting the most aggressive tumors Mapper identifies are also the most treatment-responsive. A global model confirmed the High Risk tier retained a significantly elevated hazard ratio even after adjusting for age, stage, treatment, proliferation score, and subtype, confirming that Mapper-derived topology carries prognostic information beyond conventional clinical variables.\\
\noindent \textbf{Biological Validation via Proliferation Score:} To verify biological validity, mean expression of the six proliferation marker genes (\textit{MKI67}, \textit{CCNB1}, \textit{MYBL2}, \textit{BIRC5}, \textit{UBE2C}, \textit{RRM2}) was computed per tier, revealing a strict monotone gradient: High Risk scored highest (0.1552), Mid Risk intermediate ($-0.1595$), and Low Risk lowest ($-0.1634$). A one-way ANOVA confirmed significant inter-tier differences ($p = 0.000158$), showing that the purely topological risk separation aligns with an established biological axis despite the graph never having access to proliferation measurements.

\subsection{Assignment of Validation Patients and Kaplan--Meier Analysis}

Since validation samples were not in the training graph, their risk tiers were assigned using a correlation-based transfer approach: Pearson correlation was computed between each validation sample and every training node's average expression vector, and the sample took the tier of the node with the highest correlation if it exceeded $0.65$. Otherwise, a Positive Borda Rank was used: negative or zero correlations were ignored, the remaining nodes were ranked by correlation, each tier received a weighted score (size $\times$ rank weight), and the sample was assigned to the tier with the highest score; samples with no positive correlation were labeled Unclassified.

\begin{figure}[H]
  \centering
  \begin{subfigure}{0.475\linewidth}
    \centering
    \includegraphics[width=\linewidth]{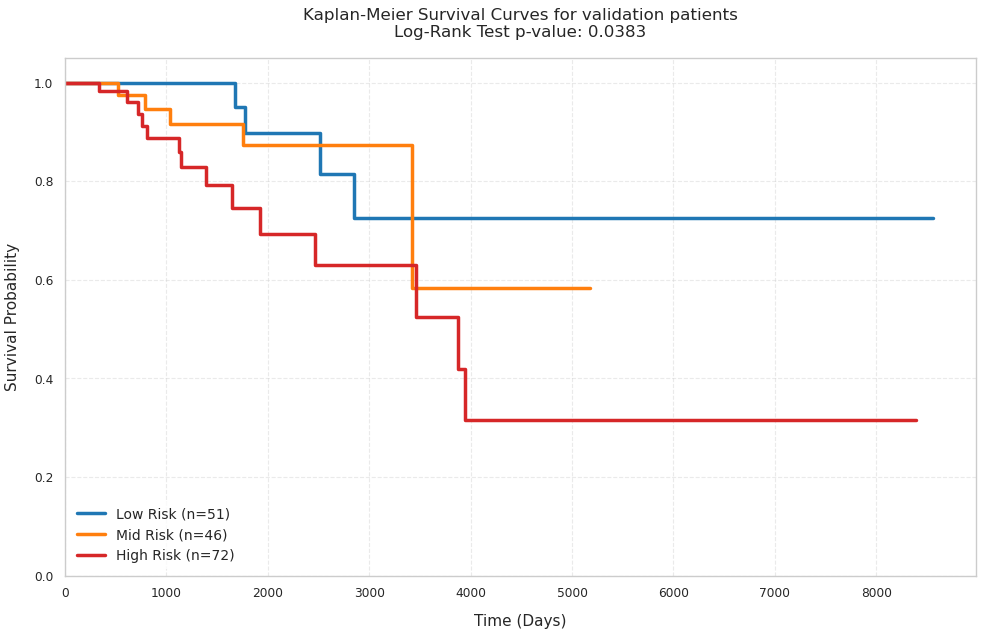}
    \caption{All three risk tiers.}
    \label{fig:km_val}
  \end{subfigure}\hfill
  \begin{subfigure}{0.475\linewidth}
    \centering
    \includegraphics[width=\linewidth]{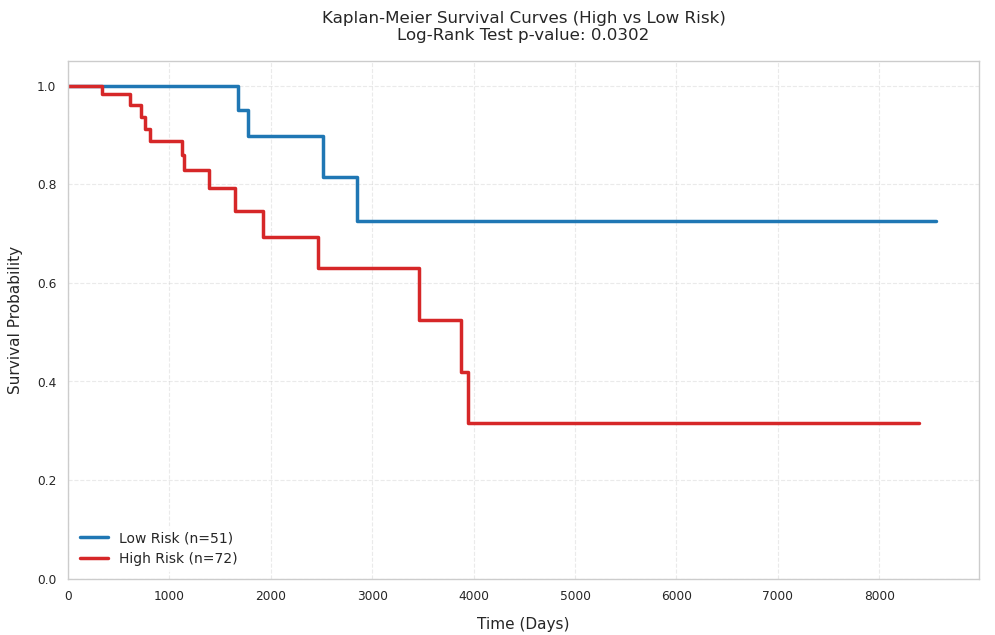}
    \caption{High Risk vs.\ Low Risk only.}
    \label{fig:km_val_hl}
  \end{subfigure}
  \caption{Kaplan--Meier survival curves for held-out validation patients. (a) All three Mapper-derived risk tiers: the risk ordering from training is preserved, and the global log-rank test remains statistically significant. (b) Pairwise comparison of the two extreme tiers confirms that the High-Low Risk contrast is the primary driver of the overall separation.}
  \label{fig:km_validation_main} 
\end{figure}

Kaplan--Meier curves for the three validation risk groups (Figure~\ref{fig:km_val}) preserved the training order: High Risk had the lowest survival, Low Risk the best, and Mid Risk intermediate, with a significant global log-rank test. A pairwise comparison of High Risk and Low Risk (Figure~\ref{fig:km_val_hl}) confirmed that the contrast between the two extreme tiers drove the overall significance.

\subsubsection{Cox Proportional Hazards Regression on the Validation Set}
A final Cox model on the validation cohort, adjusting for age, stage, and treatment, confirmed that High Risk patients faced substantially elevated mortality relative to Low Risk, with Mid Risk intermediate; age and stage remained independently significant, as in training, and treatment retained a protective effect most visible in the High Risk tier, mirroring the training cohort's topology-therapy interaction. Pairwise log-rank tests confirmed all three tier contrasts were statistically significant. Clinically, this confirms that Mapper-derived risk tiers generalize beyond the patients used to build the graph: a tier assigned from gene expression alone carries real, independently verified survival information, supporting its use as a generalizable risk-stratification tool.

%\section{Concook7lusions}

\section{Conclusions}

A Mapper-based survival analysis of breast cancer patients was conducted to investigate whether a patient's position within a gene expression network carries survival information beyond conventional PAM50 classification, and whether this topological structure generalizes to unseen patients. The dual-Mapper framework of Rostami et al.~\cite{rostami2025tda} was reproduced on TCGA-BRCA and extended with node-level survival analysis, discordant survivor identification, and train-test validation using Cox proportional hazards modeling.
Mapper-derived risk tiers, built from gene expression alone, preserved their risk ordering in held-out patients and remained significant after adjusting for age, stage, and treatment. Discordant survivor analysis identified patients whose outcomes diverged from their expected PAM50-associated prognosis, concentrated at topological subtype boundaries and associated with a cross-subtype gene signature reversal. These findings suggest Mapper topology captures candidate treatment-associated biology not captured by PAM50 classification alone, supporting topological data analysis as a complementary framework to existing subtyping.
Limitations include reliance on a single TCGA cohort, sensitivity of Mapper to lens and cover parameters, the loss of independent significance for landmark distance after adjusting for age and stage, and the absence of external validation. Future work should assess associations between treatment and survival across topological risk groups, validate these findings externally, and explore outcome-aware lenses~\cite{fores2022progression}. External validation and benchmark comparison remain necessary before clinical translation.
\newpage
\bibliographystyle{unsrtnat}
\bibliography{references}
\end{document}